\documentclass[twocolumn,aps,pra,superscriptaddress,showpacs,tightenlines]{revtex4-1}
\usepackage{amsmath}
\usepackage{amsfonts}
\usepackage{graphicx}
\usepackage{epsfig}
\usepackage{color}
\usepackage{hyperref}
\hypersetup{
    colorlinks=true, 
    linktoc=all,     
    linkcolor=blue,  
    citecolor=blue}

\begin{document}

\title{Nonreciprocal State Conversion between Microwave and Optical Photons}

\author{Lin Tian}

\affiliation{University of California, Merced, 5200 North Lake Road, Merced, California 95343, USA}

\author{Zhen Li}

\affiliation{Department of Applied Physics, School of Science, Xi'an Jiaotong University, Xi'an 710049, People's Republic of China}
\affiliation{University of California, Merced, 5200 North Lake Road, Merced, California 95343, USA}

\begin{abstract}
Optoelectromechanical quantum interfaces can be utilized to connect systems with distinctively different frequencies in hybrid quantum networks. Here we present a scheme of nonreciprocal quantum state conversion between microwave and optical photons via an optoelectromechanical interface. By introducing an auxiliary cavity and manipulating the phase differences between the linearized light-matter couplings, uni-directional state transmission that is immune to mechanical noise can be achieved. This interface can function as an isolator, a circulator, and a two-way switch that routes the input state to a designated output channel. We show that under a generalized impedance matching condition, the state conversion can prevent thermal fluctuations of the mechanical mode from propagating to the cavity outputs and reach high fidelity. The realization of this scheme is also discussed. 
\end{abstract}

\maketitle

\section{Introduction\label{sec:intro}}

The past decade has witnessed enormous progress in the study of opto- and electro-mechanical systems in the quantum limit with experimental milestones such as the realization of cavity cooling to the mechanical ground state~\cite{optomech_rev1, optomech_rev3, cooling1, cooling2, cooling3}. Mechanical resonators can be coupled to a broad variety of electronic, atomic, and photonic systems, ranging from acoustic to optical frequencies~\cite{optomech_rev2}. Optoelectromechanical systems can hence serve as an interface to bridge devices with distinctively different frequencies in hybrid quantum networks and advance the development of scalable quantum processors~\cite{interface1, interface2, hybrid1, hybrid2}. Bi-directional state conversion and entanglement generation between microwave and optical photons have been realized via optoelectromechanical interfaces~\cite{interface_rev1, interface_exp1, interface_exp2, interface_exp3, interface_exp4, interface_exp5, interface_exp6}. And mechanical dark mode that can facilitate high-fidelity state transfer has been demonstrated~\cite{darkmode1, darkmode2, darkmode3}.

Nonreciprocal devices, such as circulators and isolators, are of crucial importance in the realization of noiseless and lossless quantum networks~\cite{nonreciprocal_rev1, nonreciprocal_rev2, nonreciprocal_rev3, nonreciprocal1}. In these devices, the transmission of information is not symmetric. For example, quantum states at the input of one mode can be transmitted to the output of another mode, but not vice versa. Various effects have been exploited to implement nonreciprocal devices, including the Faraday rotation effect in magneto-optical crystals~\cite{nonreciprocal_rev1, nonreciprocal2, nonreciprocal3}, angular momentum biasing in photonic or acoustic systems~\cite{nonreciprocal4, nonreciprocal5, nonreciprocal6}, optical nonlinearity~\cite{nonreciprocal7}, and the Hall effect~\cite{nonreciprocal8}. Nonreciprocity in topological photonic devices has been implemented by generating effective magnetic fields and gauge phases with time-modulated parameters~\cite{gaugephase1, gaugephase2}, and similar approaches have been studied in quantum information applications~\cite{gaugephase3, gaugephase4, gaugephase5}. Isolators and circulators have been realized in microwave devices via parametric pumping of system parameters~\cite{nonreciprocal1, microwave1, microwave2}, and a graphic method was recently developed to facilitate the design of these devices~\cite{microwave3}. In several works~\cite{optomech1, optomech2, optomech3}, opto- and electro-mechanical systems were studied for uni-directional transmission of photon states. More recently, it was shown that nonreciprocal state conversion between directly-coupled cavities can be achieved by controlling the relative phases of the couplings in an optomechanical system~\cite{Li1}. It has also been demonstrated that nonreciprocal state conversion can be realized via quantum reservoir engineering~\cite{Clerk1,Clerk2}. In \cite{Li2}, unidirectional state transfer between microwave and optical photons via two mechanical resonators was studied; however, noise in the strongly-damped mechanical resonator can be transmitted to the cavity outputs and propagate to other parts of the quantum network. In practice, coupling between subsystems in different frequency ranges could induce serious damage to the quantum coherence of the system. Hence, despite the previous efforts, it is still challenging to implement nonreciprocal quantum interface that connects distinctively different frequencies and is robust against mechanical noise. 

Here we present a scheme of nonreciprocal state conversion between microwave and optical photons via an optoelectromechanical quantum interface, where mechanical noise can be prevented from propagating into the cavity outputs. In our system, no direct coupling exists between the microwave and the optical cavities. Instead, an auxiliary cavity is used to control the direction of the state flow. We find that by manipulating the phase differences and by adjusting the magnitudes of the linearized couplings to satisfy a generalized impedance matching condition, nearly perfect nonreciprocal state conversion can be achieved. The interface can function not only as an isolator and a circulator, but also as a two-way switch that routes the input signal as demanded. This scheme is closely related to the engineering of effective magnetic flux in photonic and atomic systems~\cite{gaugephase2, gaugephase3, gaugephase4}. During the conversion, thermal fluctuations are largely confined within the mechanical mode, which ensures high fidelity for the output state at finite temperature. Our scheme can be realized with current experimental technology~\cite{interface_exp1, interface_exp2, interface_exp3, interface_exp4, interface_exp5}, and it provides a practical approach to achieving nonreciprocal conversion of quantum information between microwave and optical frequencies.

This paper is organized as follows. In Sec.~\ref{sec:model}, we present the model, the Langevin equation, and the transmission matrix between the input and output operators of the nonreciprocal interface. We then study state conversion via this interface at frequency $\omega=0$ and derive the optimal condition for high-fidelity state transfer in Sec.~\ref{sec:nsc}. The effect of the mechanical noise is also discussed in this section. In Sec.~\ref{sec:halfwidth}, we analyze the transmission matrix elements at frequency $\omega$ and estimate the halfwidth of the transmission spectrum for high-fidelity nonreciprocal state conversion. In Sec.~\ref{sec:weakcoupling}, we study this system in the weak-coupling limit using the adiabatic elimination technique. The experimental realization of this scheme and practical parameters are discussed in Sec.~\ref{sec:realization}. Conclusions are given in Sec.~\ref{sec:conclusions}.

\section{Model\label{sec:model}}

\begin{figure}
\includegraphics[width=8cm,clip]{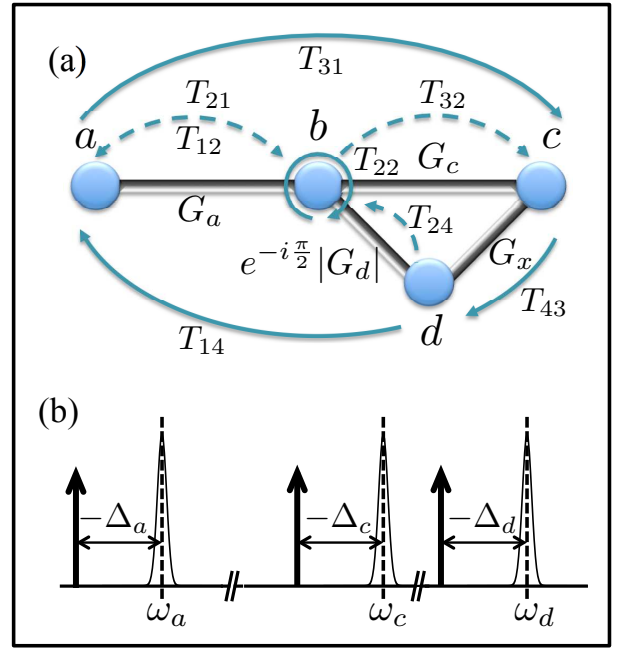}
\caption{(a) Schematic of a nonreciprocal optoelectromechanical interface with cavity modes $a, c, d$ and mechanical mode $b$. The thick bars indicate the couplings $G_{\alpha}$ ($\alpha=a, c, d$) and $G_{x}$. The solid (dashed) arrows correspond to transmission matrix elements of order $1$ ($\sqrt{\gamma_{m}/\Gamma_{\alpha}}$). (b) Driving frequencies indicated by vertical arrows together with their corresponding cavity resonances $\omega_{\alpha}$ and detunings $\Delta_{\alpha}$.}
\label{fig1} 
\end{figure}

Our system is an optoelectromechanical quantum interface composed of three cavity modes $a, c, d$ and one mechanical mode $b$, as illustrated in Fig.~\ref{fig1}(a). The auxiliary cavity $d$ is introduced to facilitate nonreciprocal state conversion between cavities $a$ and $c$, and it is directly coupled to cavity $c$. Cavity $a$ has distinctively different frequency from that of $c$ and $d$, e.g., $a$ can be a microwave resonator and $c, d$ are optical cavities, or vice versa. A direct coupling between a superconducting microwave resonator and an optical cavity can excite quasiparticles in the superconductor, which destroys the coherence of the microwave resonator~\cite{hybrid1}. In our setup, no direct coupling exists between $a$ and cavities $c, d$. All three cavities are coupled to the mechanical resonator via radiation-pressure interaction~\cite{Law1995}. By applying strong driving on the cavities, as shown in Fig.~\ref{fig1}(b), the light-matter interaction can be linearized and the total Hamiltonian becomes $\hat{H}_{t} = \hat{H}_{0} + \hat{H}_{int}$ in the rotating frame of the driving fields~\cite{Vitali2008}. The uncoupled Hamiltonian is ($\hbar=1$)
\begin{equation}
\hat{H}_{0}=\sum_{\alpha}\left(-\Delta_{\alpha}\right)\hat{\alpha}^{\dagger}\hat{\alpha}+\omega_{m}\hat{b}^{\dagger}\hat{b}, \label{eq:H0}
\end{equation} 
where $\hat{\alpha}$ ($\hat{b}$) is the annihilation operator of cavity mode $\alpha$ with $\alpha=a, c, d$ (mechanical mode), $\Delta_{\alpha}$ is the detuning of the cavity under the driving field, and $\omega_{m}$ is the frequency of the mechanical mode. We choose $-\Delta_{\alpha}=\omega_{m}\gg |G_{\alpha}|$ with $G_{\alpha}$ being the linearized coupling between cavity $\alpha$ and mechanical mode $b$. The magnitude and phase of $G_{\alpha}$ can be controlled by adjusting the driving field. Here the driving fields are assumed to be in the linear regime, where the nonlinear terms in radiation-pressure interaction are negligible. Under the rotating-wave approximation, the fast-oscillating counter-rotating terms in the interaction can be omitted and the interaction Hamiltonian is simplified as
\begin{equation}
\hat{H}_{int} = \sum_{\alpha}\left(G_{\alpha}\hat{\alpha}^{\dag}\hat{b}+G_{\alpha}^{\star}\hat{b}^{\dagger}\hat{\alpha}\right)+G_{x}\left(\hat{c}^{\dagger}\hat{d}+\hat{d}^{\dag}\hat{c}\right),\label{eq:Hint}
\end{equation}
where $G_{x}$ is the photon hopping between cavities $c$ and $d$~\cite{note1}. Details of the derivation of this Hamiltonian are given in Appendix~\ref{ssec:Ht}. The $G_{\alpha}$ and $G_{\alpha}^{\star}$ terms in (\ref{eq:Hint}) generate beam-splitter operations that are essential to cavity cooling and quantum state conversion. 

The cavities have damping rates $\kappa_{\alpha}$, which are assumed to be due to external dissipation only, with the cavity input fields $\hat{\alpha}_{in}(t)$. Similarly, the mechanical mode has damping rate $\gamma_{m}$ with the mechanical input operator $\hat{b}_{in}(t)$ satisfying $\langle  \hat{b}_{in}^{\dag}(t) \hat{b}_{in}(t^{\prime})\rangle =n_{\rm th}\delta(t-t^{\prime})$, where $n_{{\rm th}}$ is the thermal phonon occupation number at finite temperature. In Appendix~\ref{ssec:stability}, we show that this system always satisfies the stability criterion~\cite{DeJesusPRA1987}.

We define a vector $\hat{v}=[\hat{a},\hat{b},\hat{c},\hat{d}]^{\text{T}}$ in terms of the annihilation operators of the system modes. The Langevin equation of $\hat{v}$ in the interaction picture of Hamiltonian $\hat{H}_{0}$ can be written as 
\begin{equation}
id\hat{v}/dt=M\hat{v}+i\sqrt{K}\hat{v}_{in}\label{eq:langevin}
\end{equation}
with the matrix  
\begin{equation}
M=\left(\begin{array}{cccc}
-i\kappa_{a}/2 & G_{a} & 0 & 0\\
G_{a}^{\star} & -i\gamma_{m}/2 & G_{c} & G_{d}\\
0 & G_{c}^{\star} & -i\kappa_{c}/2 & G_{x}\\
0 & G_{d}^{\star} & G_{x} & -i\kappa_{d}/2
\end{array}\right),\label{eq:M}
\end{equation}
the diagonal matrix $K=\text{Diag}[\kappa_{a}, \gamma_{m}, \kappa_{c}, \kappa_{d}]$, and the input vector $\hat{v}_{in}=[\hat{a}_{in},\hat{b}_{in},\hat{c}_{in},\hat{d}_{in}]^{\text{T}}$~\cite{darkmode2}. For the output fields, we let $\hat{v}_{out}=[\hat{a}_{out},\hat{b}_{out},\hat{c}_{out},\hat{d}_{out}]^{\text{T}}$. Following the convention in \cite{darkmode2} and using the input-output theorem~\cite{inputoutput}, we have $\hat{v}_{out}=\hat{v}_{in}-\sqrt{K} \hat{v}$. The output states can then be obtained. In (\ref{eq:M}), the Langevin equation of the annihilation operators is decoupled from that of the creation operators, as the interaction Hamiltonian (\ref{eq:Hint}) only contains beam-splitter operations. The Langevin equation of the creation operators is
\begin{equation}
id\hat{v}^{\dag}dt=-\hat{v}^{\dag}M^{\dag}+i\hat{v}_{in}^{\dag}\sqrt{K},\label{eq:langevindag}
\end{equation}
where $\hat{v}^{\dag}=[\hat{a}^{\dag},\hat{b}^{\dag},\hat{c}^{\dag},\hat{d}^{\dag}]$ and $\hat{v}_{in}^{\dag}=[\hat{a}_{in}^{\dag},\hat{b}_{in}^{\dag},\hat{c}_{in}^{\dag},\hat{d}^{\dag}_{in}]$. 

Using the transformation $\hat{o}(t)=\int d\omega e^{-i\omega t} \hat{o}(\omega) /2\pi$ for an arbitrary operator $\hat{o}$, Eq.~(\ref{eq:langevin}) can be converted to the frequency domain with $\hat{v}(\omega)=i(\omega I-M)^{-1}\sqrt{K}\hat{v}_{in}(\omega)$, where $I$ is the $4\times4$ identity matrix. With the input-output theorem~\cite{inputoutput}, we derive $\hat{v}_{out}=T(\omega)\hat{v}_{in}$, where 
\begin{equation}
T(\omega)=I-i\sqrt{K}(\omega I-M)^{-1}\sqrt{K}.\label{eq:wtT}
\end{equation}
is the transmission matrix of this interface. It can be shown that $T(\omega)$ is a unitary matrix.

\section{Nonreciprocal state conversion\label{sec:nsc}}

Without loss of generality, we assume that the couplings $G_{a, c, x}$ are positive numbers and $G_{d}$ carries a nontrivial phase. In this section, we consider the state conversion of an input field in resonance with the cavity frequency, i.e., $\omega=0$ in the interaction picture. To achieve nonreciprocity, it requires that the transmission matrix element $T_{13}$ describing state conversion from $c$ to $a$ be zero and $T_{31}$ describing state conversion from $a$ to $c$ approach unity. With (\ref{eq:wtT}), we find 
\begin{equation}
\frac{T_{13}}{T_{31}}=\frac{G_{c}-2iG_{d}G_{x}/\kappa_{d}}{G_{c}-2iG_{d}^{*}G_{x}/\kappa_{d}}.\label{eq:T13T31}
\end{equation}
The nonreciprocal conditions are $G_{d}=e^{-i\pi/2}|G_{d}|$ with a nontrivial phase $(-\pi/2)$ and $|G_{d}|=G_{c}\kappa_{d}/2G_{x}$. The disappearance of $T_{13}$ results from the destructive quantum interference between two possible paths for the state conversion. In one path, the input state of cavity $c$ is transferred to the output of cavity $a$ along $c\rightarrow b\rightarrow a$ with a transmission amplitude proportional to $G_{c}$. In the other path, the state transfer is facilitated by the coupling $G_{x}$ and is along $c\rightarrow d\rightarrow b\rightarrow a$ with an amplitude proportional to $-2iG_{d} G_{x}/i\kappa_{d}$. By choosing a $(-\pi/2)$ phase for $G_{d}$, the amplitudes of these two paths cancel each other. In contrast, the two paths for the state conversion from $a$ to $c$ have transmission amplitudes proportional to $G_{c}$ and $-2iG_{d}^{\star} G_{x}/i\kappa_{d}$, respectively. Because $G_{d}^{\star}=-G_{d}$, the two paths interfere constructively to enhance the matrix element $T_{31}$. The coupling $G_{x}$ together with the coupling $G_{d}$ provides an indirect route for state conversion between modes $a$ and $c$ that interferes with the direct transmission between $a$ and $c$ via the coupling $G_{c}$. This interference is crucial for achieving nonreciprocity. 

To prevent loss of the input photon to other modes in the interface, it requires that $|T_{i1}/T_{31}|\ll1$ ($i=1, 2, 4$). By choosing $G_{x}=(\sqrt{\kappa_{c}\kappa_{d}})/2$, we have $|T_{41}/T_{31}|=0$. Together with the nonreciprocal condition discussed above, $|G_{d}|=G_{c}\sqrt{\kappa_{d}/\kappa_{c}}$, which is equivalent to the impedance matching condition $G_{c}^{2}/\kappa_{c}=G_{d}^{2}/\kappa_{d}$ between cavities $c$ and $d$~\cite{darkmode1, darkmode2}. At weak mechanical damping $\gamma_{m}\ll 4G_{c}^{2}/\kappa_{c}$, $|T_{21}/T_{31}|=\sqrt{\kappa_{c}\gamma_{m}}/2G_{c}\ll1$. 

\begin{figure}
\includegraphics[width=8cm,clip]{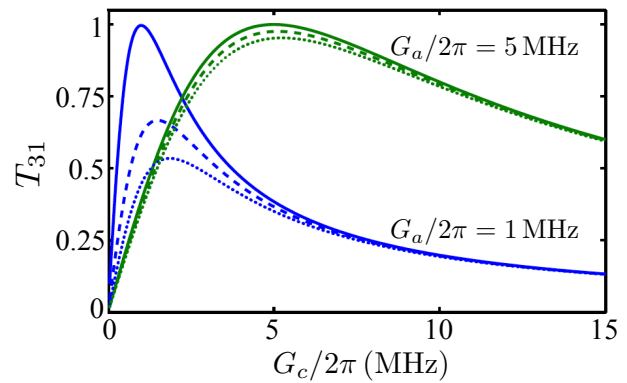}
\caption{The transmission matrix element $T_{31}$ vs the coupling $G_{c}$ under the impedance matching condition $G_{c}^{2}/\kappa_{c}=G_{d}^{2}/\kappa_{d}$ and the nonreciprocal conditions. Here $\kappa_{a,c,d}/2\pi=5\,\textrm{MHz}$ and $G_{a}/2\pi=(1, 5)\,\textrm{MHz}$. The solid, dashed, and dotted curves correspond to $\gamma_{m}/2\pi=(0.005, 1, 2)\,\textrm{MHz}$, respectively.}
\label{fig2} 
\end{figure}

Under the above nonreciprocal and low-loss conditions, it can be obtained that
\begin{equation}
T_{31}=\frac{8G_{c}G_{a}\sqrt{\kappa_{a}\kappa_{c}}}{4G_{a}^{2}\kappa_{c}+4G_{c}^{2}\kappa_{a}+\kappa_{a}\kappa_{c}\gamma_{m}}.\label{eq:T31}
\end{equation}
In Fig.~\ref{fig2}, we plot this transmission matrix element as a function of the coupling constant $G_{c}$ at various $G_{a}$ and $\gamma_{m}$ values. Here we choose the cavity damping rates to be $\kappa_{a,c}/2\pi=5\,\textrm{MHz}$. Practical parameters of the damping rates and the coupling constants will be discussed in Sec.~\ref{sec:realization}. It can be shown that for given values of $G_{a}$ and $\gamma_{m}$, maximum transmission can be reached at the optimal value of $G_{c}=\sqrt{\widetilde{\gamma}_{m}\kappa_{c}}/2$ with $\widetilde{\gamma}_{m}=\Gamma_{a}+\gamma_{m}$, where $\Gamma_{\alpha}=4G_{\alpha}^{2}/\kappa_{\alpha}$ for each cavity mode. In the weak-coupling limit with $G_{\alpha}\ll\kappa_{\alpha}$, $\Gamma_{\alpha}$ corresponds to the cooling rate that cavity $\alpha$ exerts on the mechanical mode~\cite{optomech_rev1}. At the optimal coupling, the transmission matrix element can be written as $T_{31}=\sqrt{\Gamma_{a}/\widetilde{\gamma}_{m}}$. The transmission can hence be enhanced by increasing the power of the driving fields. With $\gamma_{m}\ll \Gamma_{a}$, the maximum transmission $T_{31}\approx1-\gamma_{m}/2\Gamma_{a}$, and high-fidelity nonreciprocal state conversion can be achieved. Furthermore, this optimal condition is equivalent to $\Gamma_{a}\approx \Gamma_{c}=\Gamma_{d}$, which is a generalized impedance matching condition for these three cavities.

The optimal condition also ensures that the output state is robust against mechanical noise and unwanted photon fields from other cavities. With $T_{31}\rightarrow 1$, $|T_{3j}/T_{31}|\rightarrow 0$ ($j=2, 3, 4$), as the transmission matrix is unitary. The full transmission matrix is 
\begin{equation}
T=\left(\begin{array}{cccc}
-\gamma_{m}/\Gamma_{a} & i\sqrt{\gamma_{m}/\Gamma_{a}} & 0 & -i(1-\gamma_{m}/2\Gamma_{a})\\
i\sqrt{\gamma_{m}/\Gamma_{a}} & 1-\gamma_{m}/\Gamma_{a} & 0 & \sqrt{\gamma_{m}/\Gamma_{a}}\\
1-\gamma_{m}/2\Gamma_{a} & i\sqrt{\gamma_{m}/\Gamma_{a}} & 0 & 0\\
0 & 0 & i & 0
\end{array}\right),\label{eq:T}
\end{equation}
where we only keep the lowest order $(\gamma_{m}/\Gamma_{a})$ term in each matrix element. Using this matrix, the output of cavity $c$ can be written as 
\begin{equation}
\hat{c}_{out}=(1-\gamma_{m}/2\Gamma_{a})\hat{a}_{in}+i\sqrt{\gamma_{m}/\Gamma_{a}}\hat{b}_{in},\label{eq:cout}
\end{equation}
which is dominated by the input field $\hat{a}_{in}$. The contribution of the mechanical noise $\hat{b}_{in}$ is suppressed by a factor $\sqrt{\gamma_{m}/\Gamma_{a}}$, which makes it possible to achieve high-fidelity nonreciprocal state conversion at finite temperature. The output field of cavity $a$ is 
\begin{equation}
\hat{a}_{out} =i\sqrt{\gamma_{m}/\Gamma_{a}}\hat{b}_{in}-i(1-\gamma_{m}/2\Gamma_{a})\hat{d}_{in},\label{eq:aout}
\end{equation}
where we omit the small term $(-\gamma_{m}/\Gamma_{a})\hat{a}_{in}$. This output field contains no contribution from the input field $\hat{c}_{in}$, clearly demonstrating the nonreciprocity of this scheme. Instead, it is dominated by the input field $\hat{d}_{in}$ with the mechanical noise suppressed by the factor $\sqrt{\gamma_{m}/\Gamma_{a}}$. For the output field of cavity $d$, $\hat{d}_{out} = i \hat{c}_{in}$, i.e., the input of cavity $c$ is fully transferred to the output of cavity $d$. Meanwhile, the mechanical output $\hat{b}_{out} \approx \hat{b}_{in}$ to leading order with the mechanical noise mainly confined in the mechanical mode. 

For an input field at the single-photon level, it requires that the cooperativity $\Gamma_{a}/\gamma_{m}n_{\rm th}>1$ for the state conversion to be robust against thermal fluctuations. With practical parameters~\cite{cooling1, cooling2, cooling3}, $\gamma_m/\Gamma_a \sim 10^{-6}$ can be reached. High-fidelity state conversion is hence possible for $n_{\rm th} < 10^{6}$. It also shows that the mechanical noise transferred to the cavities can be suppressed by three orders of magnitude and will not spread significantly to the cavity modes. 

This optoeletromechanical interface functions as a circulator under the optimal condition. Quantum states are transmitted with high fidelity along the route $a\rightarrow c\rightarrow d\rightarrow a$~\cite{nonreciprocal_rev2}. Meanwhile, by flipping the phase of the coupling $G_{d}$ from $(-\pi/2)$ to $\pi/2$, i.e., $G_{d}=e^{i\pi/2}|G_{d}|$, the cavity inputs are transmitted coherently along the opposite direction $c\rightarrow a\rightarrow d\rightarrow c$, which can be illustrated by reversing the directions of all the arrows in Fig.~\ref{fig1}(a) and simultaneously changing the labels from $T_{ij}$ to $T_{ji}$. Furthermore, this interface can be utilized as a two-way switch. By selecting the phase of $G_{d}$ as $(-\pi/2)$ (or as $\pi/2$), the input state $\hat{a}_{in}$ can be routed to the output of cavity $c$ (or $d$) on-demand. The auxiliary cavity $d$ plays an essential role in this scheme. Without cavity $d$, this interface is a standard three-mode system with cavities $a, c$ coupled to mechanical mode $b$~\cite{interface_rev1, interface_exp1, interface_exp2, interface_exp3, interface_exp4, interface_exp5}, where  $T_{13} \equiv T_{31}$ for input fields at arbitrary frequency and state conversion is always symmetric~\cite{darkmode2}.

\section{Conversion halfwidth\label{sec:halfwidth}}

\begin{figure}
\includegraphics[width=8cm,clip]{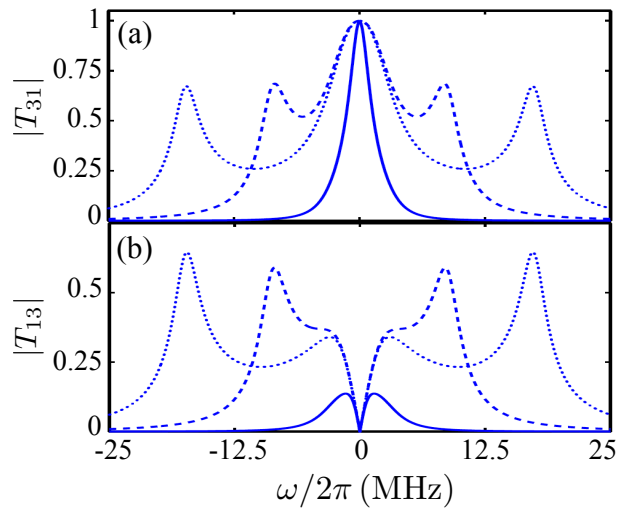}
\caption{The transmission matrix elements (a) $|T_{31}|$ and (b) $|T_{13}|$ vs the frequency of the input field $\omega$ under the optimal transmission conditions. Here $\kappa_{a,c,d}/2\pi=5\,\textrm{MHz}$ and $\gamma_{m}/2\pi=0.005\,\textrm{MHz}$. The solid, dashed, and dotted curves are for $G_{a}/2\pi=(1, 5, 10)\,\textrm{MHz}$, respectively.}
\label{fig3}
\end{figure}

With Eq.~(\ref{eq:wtT}), we study the frequency dependence of the nonreciprocal state conversion. In Fig.~\ref{fig3}(a), the matrix element $|T_{31}|$ is plotted versus the frequency of the input field $\hat{a}_{in}$ under the optimal condition. Here $\omega=0$ corresponds to the resonant frequency of the respective cavities. It can be seen that $|T_{31}|$ has a finite halfwidth near $\omega=0$. Assume that the cavity damping rates $\kappa_{\alpha}$ are all of the same order of magnitude and the couplings $G_{\alpha}$ are all of the same order of magnitude. To the first order of $\omega$, the denominator of $T_{31}$ is $-4\Gamma_{\alpha} O(\kappa_{\alpha}^{3}) + 4i[O(\kappa_{\alpha}^{3}) + \Gamma_{\alpha} O(\kappa_{\alpha}^{2})]\omega$. With this expression, we find that the halfwidth of the transmission spectrum $\Delta\omega\sim\min(\Gamma_{\alpha}, \kappa_{\alpha})$. For $|G_{\alpha}|<\kappa_{\alpha}/2$, $\Delta\omega\sim \Gamma_{\alpha}$; and for $|G_{\alpha}|>\kappa_{\alpha}/2$, $\Delta\omega\sim\kappa_{\alpha}$. This analysis reveals that the halfwidth of the transmission spectrum is upper-bounded by the damping rates of the cavities. 

For $|G_{\alpha}|>\kappa_{\alpha}$, two side peaks appear in the transmission spectrum. At the position of these side peaks, $|T_{13}|$ for state conversion from $c$ to $a$ also becomes significant, as shown in Fig.~\ref{fig3}(b). The state conversion at these frequencies is hence not unidirectional. In contrast, near $\omega=0$, $|T_{13}|$ approaches zero with a finite halfwidth, which ensures that the input field of cavity $c$ is prevented from entering cavity $a$.

\section{Weak-coupling limit\label{sec:weakcoupling}}

In the weak-coupling limit of $|G_{a}|\ll \kappa_{a}$, we can apply the adiabatic elimination technique to study this system. We set $d\hat{a}/dt=0$ in the Langevin equation of operator $\hat{a}$ and derive~\cite{interface_rev1}
\begin{equation}
\hat{a}=\left(-2iG_{a}\hat{b}+2\sqrt{\kappa_{a}}\hat{a}_{in}\right)/\kappa_{a}.\label{eq:a}
\end{equation}
Substituting (\ref{eq:a}) into the Langevin equation of operator $\hat{b}$, we have
\begin{equation}
id\hat{b}/dt = -i(\widetilde{\gamma}_{m}/2)\hat{b}+G_{c}\hat{c}+G_{d}\hat{d}+i\sqrt{\widetilde{\gamma}_{m}}\hat{b}^{\prime}_{in}\label{eq:dbdt}
\end{equation}
with $\hat{b}^{\prime}_{in}=\sqrt{\gamma_{m}/\widetilde{\gamma}_{m}}\hat{b}_{in}-i\sqrt{\Gamma_{a}/\widetilde{\gamma}_{m}}\hat{a}_{in}$ being the effective input operator of the mechanical mode and $\widetilde{\gamma}_{m}$ being the effective damping rate of the mechanical mode. Under the condition $\Gamma_{a}/\gamma_{m}n_{\rm th}\gg1$, $\hat{b}^{\prime}_{in}\approx -i \hat{a}_{in}$. With (\ref{eq:dbdt}), we find that modes $b, c, d$ form a closed loop. High-fidelity nonreciprocal state conversion from $\hat{b}^{\prime}_{in}$ to the output $\hat{c}_{out}$ can be achieved under the optimal condition derived in Sec.~\ref{sec:nsc}. The loop also acts like a circulator for the input states of these three modes. This result agrees with the exact solution in Sec.~\ref{sec:nsc}.

\section{Realization\label{sec:realization}}

Optoelectromechanical interfaces that connect microwave and optical systems have been realized in several experiments~\cite{interface_rev1, interface_exp1, interface_exp2, interface_exp3, interface_exp4, interface_exp5, interface_exp6}. Such an interface usually includes a microwave cavity or field, an optical cavity or field, and a mechanical resonator. To implement our scheme, an auxiliary cavity coupled to either the microwave or the optical cavity with time-dependent interaction needs to be added to the interface~\cite{note1}. When $c, d$ are microwave cavities, their interaction can be generated by coupling both cavities to an inductive loop~\cite{Tian2008}. By modulating the magnetic flux in the loop, the interaction in (\ref{eq:Htp}) can be realized with its magnitude reaching $1-100\,{\rm MHz}$. When $c, d$ are optical cavities, time-dependent coupling can be generated by connecting the cavities to other cavity modes or waveguides~\cite{gaugephase2, Sato2012, Cho2008}.  For cavities in both microwave and optical regimes, time-dependent interaction can also be generated by coupling the cavities to a quantum two-level system with tunable energy splitting, such as qubit and defect~\cite{Baust2015, Sharypov2012}. 

For practical parameters, consider a mechanical mode with frequency $\omega_{m}/2\pi = 100\,{\rm MHz}$ and $\gamma_{m}/2\pi=100\,{\rm Hz}$ (quality factor $Q_{m}=10^{6}$). For both microwave and optical cavities, the damping rate can be $\kappa_{\alpha}/2\pi=1-10\,{\rm MHz}$ and the coupling strength $G_{\alpha}/2\pi$ can reach a few tens of MHz~\cite{cooling1, cooling2, cooling3}. Assume, e.g., $\kappa_{\alpha}/2\pi=5\,{\rm MHz}$. Our scheme requires that $G_{x}/2\pi=2.5\,{\rm MHz}$. With $|G_{\alpha}|/2\pi=5\,{\rm MHz}$, the cooling rate is $\Gamma_{\alpha}/2\pi=20\,{\rm MHz}$, and $\Gamma_{\alpha}/\gamma_{m}=2\times10^{5}$.

In the previous sections, we neglect the effect of intrinsic cavity dissipation on the state conversion. In practice, cavity damping rate is a sum of the external damping rate $\kappa_{\alpha}^{ext}$, which describes the coupling between a cavity and its output channels, and the intrinsic damping rate $\kappa_{\alpha}^{in}$, which describes the dissipation of cavity photons in internal channels, with $\kappa_{\alpha}=\kappa_{\alpha}^{ext}+\kappa_{\alpha}^{in}$. For a finite intrinsic damping rate, the input field $\hat{a}_{in}$ transmitted to the output field $\hat{c}_{out}$ is reduced to be $\sqrt{\kappa_{a}^{ext}\kappa_{c}^{ext}/\kappa_{a}\kappa_{c}}T_{31}\hat{a}_{in}$~\cite{interface_rev1}. Meanwhile, the output field $\hat{c}_{out}$ includes small contribution from the internal noise terms, such as $\hat{a}_{in}^{(n)}$ of cavity $a$. Both effects will decrease the fidelity of the state conversion~\cite{interface_rev1}.

\section{Conclusions\label{sec:conclusions}}

To conclude, we present an optoelectromechanical quantum interface for nonreciprocal state conversion between microwave and optical photons without direct coupling between the microwave and the optical cavities. By introducing an auxiliary cavity and manipulating the phase differences between the couplings, nearly perfect nonreciprocal state conversion can be achieved. The effect of the mechanical noise is strongly suppressed under the impedance matching condition, and single-photon level state conversion with high fidelity is possible at finite temperature. This interface can serve as an isolator, a circulator, and a two-way switch for input photon states. Our scheme can be used to realize nonreciprocal transmission of quantum information in hybrid quantum networks involving distinctively different frequencies.

\section*{acknowledgments}

This work is supported by the National Science Foundation under Award No. NSF-DMR-0956064. Z. Li is also supported by a fellowship from the China Scholarship Council.

\appendix

\section{Hamiltonian in the rotating frame\label{ssec:Ht}}

The original Hamiltonian of the optoelectromechanical interface in Fig.~\ref{fig1}(a) can be written as ($\hbar=1$)
\begin{eqnarray}
\hat{H}_{t}^{\prime} &=& \sum_{\alpha}\omega_{\alpha}\hat{\alpha}^{\dagger}\hat{\alpha} + G_{\alpha}^{\prime} \hat{\alpha}^{\dag} \hat{\alpha} (\hat{b}+\hat{b}^{\dagger}) +\omega_{m}\hat{b}^{\dagger}\hat{b} \nonumber \\
&& +\sum_{\alpha}\left[\epsilon_{\alpha}(t) \hat{\alpha}^{\dagger} + \epsilon_{\alpha}^{\star}(t) \hat{\alpha} \right]  \nonumber \\
&& +G_{x}^{\prime}(t)(\hat{c}+\hat{c}^{\dagger})(\hat{d}+\hat{d}^{\dag}),\label{eq:Htp}
\end{eqnarray}
where $\omega_{\alpha}$ is the frequency of cavity mode $\hat{\alpha}$ ($\alpha=a, c, d$), $G_{\alpha}^{\prime}$ is the single-photon opto- and electro-mechanical coupling between cavity $\alpha$ and mechanical mode $b$, $\epsilon_{\alpha}(t)$ is the time-dependent driving amplitude on cavity $\alpha$, and $G_{x}^{\prime}(t)$ is the time-dependent coupling between cavities $c$ and $d$. The driving frequency $\omega_{d}^{\alpha}$ for cavity $\alpha$ is below the resonant frequency $\omega_{\alpha}$ of the respective cavity mode. By applying strong driving to the cavities, the opto- and electro-mechanical couplings can be linearized. In the rotating frame of the driving fields, the Hamiltonian has the form
\begin{eqnarray}
\hat{H}_{t} &=& \sum_{\alpha}(-\Delta_{\alpha})\hat{\alpha}^{\dagger}\hat{\alpha}+ (G_{\alpha}\hat{\alpha}^{\dag} + G_{\alpha}^{\star}\hat{\alpha}) (\hat{b}+\hat{b}^{\dagger}) +\omega_{m}\hat{b}^{\dagger}\hat{b}  \nonumber \\
&& + G_{x}^{\prime}(t)(e^{i\omega_{d}^{c}t}\hat{c}^{\dagger}+e^{-i\omega_{d}^{c}t}\hat{c})(e^{i\omega_{d}^{d}t}\hat{d}^{\dag}+e^{-i\omega_{d}^{d}t}\hat{d}),\,\,\label{eq:Htv1}
\end{eqnarray}
where $\Delta_{\alpha}=\omega_{d}^{\alpha}-\omega_{\alpha}-\delta\omega_{\alpha}$ is the detuning of cavity $\alpha$, $\delta\omega_{\alpha}$ is the small shift of the cavity resonance due to the stationary mechanical displacement, and $G_{\alpha}$ is the linearized coupling between cavity mode $\alpha$ and mechanical mode $b$. 

Let the detuning of cavity $\alpha$ be $-\Delta_{\alpha}=\omega_{m}$ and the time-dependent coupling be $G_{x}^{\prime}=2G_{x}\cos(\omega_{c}-\omega_{d})t$. Under the assumption that $\omega_{\alpha}, \omega_{d}^{\alpha}, |\Delta_{\alpha}|,\omega_{m}\gg G_{\alpha}, G_{x}$, we apply the rotating-wave approximation and neglect the fast-oscillating terms in (\ref{eq:Htv1}). The rotating-frame Hamiltonian then becomes $\hat{H}_{t} = \hat{H}_{0} + \hat{H}_{int}$ with $\hat{H}_{0}$ and $\hat{H}_{int}$ given by (\ref{eq:H0}) and (\ref{eq:Hint}), respectively.

\section{Stability\label{ssec:stability}}

The stability of this nonreciprocal quantum interface can be determined from the eigenvalues of the matrix $(-i M)$ with $M$ given by (\ref{eq:M}). Based on the Routh-Hurwitz criterion, this system is stable when the real parts of all four eigenvalues of $(-i M)$ are negative~\cite{DeJesusPRA1987}. 

It can be shown that the eigenvalues satisfy the following equation: 
\begin{equation}
\lambda^{4} + s_{3} \lambda^{3} +  s_{2} \lambda^{2} +  s_{1} \lambda + s_{0} =0. \label{eq:eigen}
\end{equation}
The coefficients $s_{i}$ ($i=0,1,2,3$) can be derived as
\begin{eqnarray}
s_{3} &=& \left(\gamma_{m} + \kappa_{a}+ \kappa_{c}+ \kappa_{d}\right)/2, \label{eq:s3} \\
s_{2} &=& \left[\kappa_{a}\kappa_{c} +\kappa_{a}\kappa_{d} +\kappa_{c}\kappa_{d}+\gamma_{m}(\kappa_{a} 
+ \kappa_{c} +\kappa_{d} )  \right] /4  \nonumber \\
&&+ \left(G_{a}^{2}+ G_{c}^2 + |G_{d}|^2  + G_{x}^{2}\right), \label{eq:s2} \\
s_{1} &=&G_{a}^{2}(\kappa_{c}+\kappa_{d})/2 + G_{c}^2 (\kappa_{a}+ \kappa_{d})/2 \nonumber \\
&& + |G_{d}|^2 (\kappa_{a} + \kappa_{c})/2 +  G_{x}^{2} (\kappa_{a} + \gamma_{m})/2 \nonumber \\
&& + \left[\kappa_{a}\kappa_{c}\kappa_{d} + (\kappa_{a}\kappa_{c}+ \kappa_{c}\kappa_{d}+ \kappa_{a}\kappa_{d})\gamma_{m}\right] /8, \label{eq:s1} \\
s_{0} &=& \left( G_{a}^{2} \kappa_{c}\kappa_{d} + G_{c}^2 \kappa_{a}\kappa_{d} + |G_{d}|^2 \kappa_{a}\kappa_{c}  + G_{x}^{2}\gamma_{m} \kappa_{a} \right)/4 \nonumber \\  
&&+ G_{a}^{2} G_{x}^{2} +    \gamma_{m}\kappa_{a}\kappa_{c}\kappa_{d} /16. \label{eq:s0}
\end{eqnarray}
The conditions for stability include: (1) all $s_{i}>0$, (2) $s_{3}s_{2}-s_{1}>0$, and (3) $s_{3}s_{2}s_{1}-s_{1}^{2}-s_{0}s_{3}^{2}>0$~\cite{DeJesusPRA1987}. All three conditions are fulfilled in our system with arbitrary parameters.


\begin{thebibliography}{99}
\bibitem{note1}In the lab frame, the coupling between cavities $c$ and $d$ is time-dependent in the form of $G_{x}^{\prime}(t)(\hat{c}+\hat{c}^{\dagger})(\hat{d}+\hat{d}^{\dag})$ with $G_{x}^{\prime}=2G_{x}\cos(\omega_{c}-\omega_{d})t$. In the rotating frame of the driving fields and with $-\Delta_{\alpha}=\omega_{m}\gg|G_{\alpha}|$, the coupling can be approximated as $G_{x}(\hat{c}^{\dagger}\hat{d}+\hat{d}^{\dag}\hat{c})$ after omitting the counter-rotating terms.
\end{thebibliography}
\end{document}